\documentclass[prd,twocolumn,showpacs,superscriptaddress,floatfix,longbibliography]{revtex4-1}
\usepackage{graphicx,amsfonts,amssymb,amsmath,xspace,dsfont}
\usepackage[colorlinks=true,citecolor=green,linkcolor=red,urlcolor=blue]{hyperref}
\usepackage{subfigure}

\usepackage{color}
\definecolor{g}{rgb}{.1,0.4,.1} 
\definecolor{b}{rgb}{0,0.2,1}
\definecolor{rouge}{rgb}{0.82,0.,0.}
\definecolor{vert}{rgb}{0.,0.82,0.}
\definecolor{orange}{rgb}{1,0.5,0.}
\definecolor{bleu}{rgb}{0.,0.,0.82}
\definecolor{m}{rgb}{0.82,0.,0.82}
\definecolor{vert2}{rgb}{0.,0.5,0.}
\definecolor{rougeclair}{rgb}{1.0,0.7,0.7}

\newcommand{\beq}{\begin{equation}}
\newcommand{\be}{\begin{equation}}
\newcommand{\beqn}{\begin{eqnarray}}
\newcommand{\eeq}{\end{equation}}
\newcommand{\ee}{\end{equation}}
\newcommand{\eeqn}{\end{eqnarray}}

\newcommand{\bem}{\begin{pmatrix}}
\newcommand{\eem}{\end{pmatrix}}

\newlength{\ldag}
\begin{document}

\title{Hofstadter butterfly of a quasicrystal}

\author{Jean-No\"el Fuchs}
\email{fuchs@lptmc.jussieu.fr}
\affiliation{Laboratoire de Physique Th\'eorique de la Mati\`ere Condens\'ee, CNRS UMR 7600, Universit\'e Pierre et Marie Curie, 4 Place Jussieu, 75252 Paris Cedex 05, France}
\affiliation{Laboratoire de Physique des Solides, CNRS UMR 8502, Universit\'e Paris-Sud, 91405 Orsay, France}

\author{Julien Vidal}
\email{vidal@lptmc.jussieu.fr}
\affiliation{Laboratoire de Physique Th\'eorique de la Mati\`ere Condens\'ee, CNRS UMR 7600, Universit\'e Pierre et Marie Curie, 4 Place Jussieu, 75252 Paris Cedex 05, France}

\begin{abstract}

The energy spectrum of a tight-binding Hamiltonian is studied for the two-dimensional quasiperiodic Rauzy tiling in a perpendicular magnetic field. This spectrum known as a Hofstadter butterfly displays a very rich pattern of bulk gaps that are labeled by four integers, instead of two  for periodic systems. The role of phason-flip disorder is also investigated in order to extract genuinely quasiperiodic properties. This geometric disorder is found to only preserve main quantum Hall gaps.

\end{abstract}


\maketitle

%
\section{Introduction}
%
Quasicrystals are nonperiodic solids that nevertheless feature long-range configurational order \cite{Shechtman84}. In reciprocal space, this order is characterized by resolution-limited Bragg peaks in the diffraction pattern whereas, in real space, it is related to the nonperiodic repetitivity of local environments. After the initial burst of interest following their discovery, quasicrystals have drawn much attention in artificial systems (with phonons \cite{Steurer07}, cold atoms \cite{Verkerk99}, photons \cite{Krauss12,Bandres16}, polaritons \cite{Tanese14}, microwaves~\cite{Vignolo16},...). These new experimental setups allow one to investigate old problems, such as the  labeling of energy gaps or the nature of wavefunctions, that were out of reach in quasicrystalline metallic alloys. 

It has long been known that a relation exists between electrons in one-dimensional quasicrystals and the integer quantum Hall effect (IQHE) for a two-dimensional electron gas. For instance, both systems share energy spectra with gaps that can be labeled with integers that are topological invariants \cite{Thouless83b,Kunz86}. A suggestive example is given by the mapping of the Hofstadter model \cite{Hofstadter76} (two-dimensional square lattice in a magnetic field) onto the Audry-Andr\'e-Harper model \cite{Harper55,Aubry80} (one-dimensional incommensurate potential). Recently, this connection has been revisited and extended to topological insulators and superconductors (see, e.g., Refs.~\cite{Krauss12,Levy15,Fulga16}).

\begin{figure}[t]
\includegraphics[width=\columnwidth]{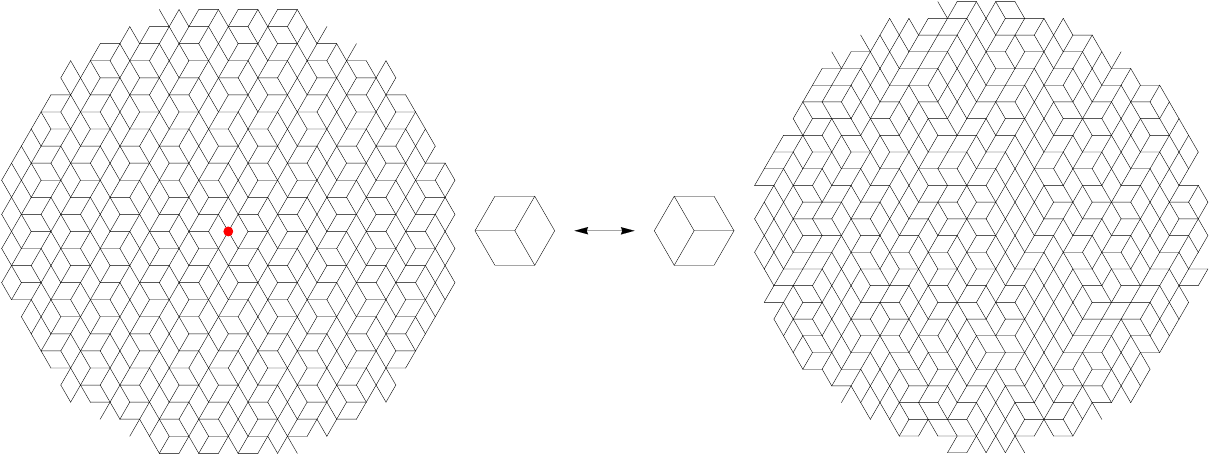}
\caption{A piece of the isometric Rauzy tiling (left) transformed into a random tiling (right) via a random sequence of phason flips (middle). The red dot indicates the center of the inversion symmetry which is broken after flips.}
\label{fig:tiling}
\end{figure}

In the present paper, we consider these two issues simultaneously by studying a two-dimensional quasicrystal in the IQHE regime. This combination  has already been addressed in the literature, but incommensurability of tile areas (in the Penrose tiling \cite{Hatakeyama89}) or edge states (in the Rauzy tiling~\cite{Vidal04,Tran15}) always prevented a complete analysis of bulk properties in the corresponding Hofstadter butterflies \cite{Hofstadter76,Satija16}. Here, we circumvent these two problems and obtain the butterfly of the Rauzy tiling using periodic boundary conditions. We find that gaps can be labeled by (four) integers related either to the IQHE or to the irrational used in the cut-and-project construction of the tiling. We also discuss the role of a structural disorder on the energy spectrum by computing the butterfly of random tilings obtained by flipping the Rauzy tiling (see Fig.~\ref{fig:tiling}).

%
%
%

%
%
\begin{figure*}[t]
\centering
\includegraphics[width=1\textwidth]{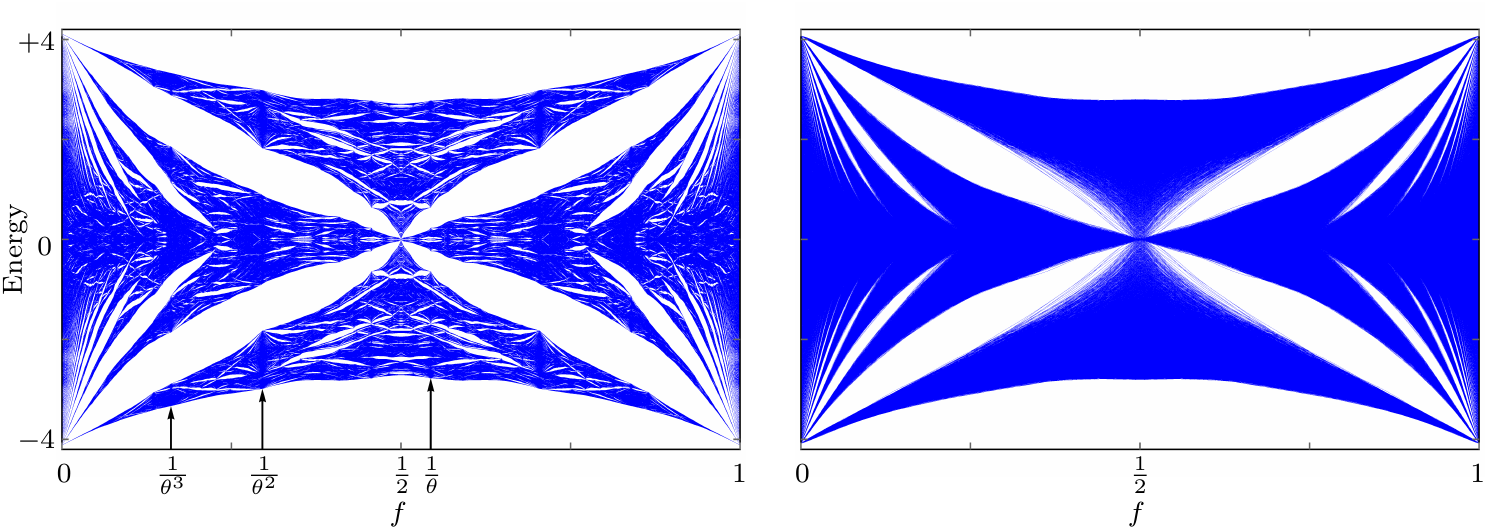}
\caption{(Color online) Hofstadter butterflies of the isometric Rauzy tiling on a torus with $R_{15}=5768$ sites (left) and of a random tiling obtained after $2.10^7$ flips (right).  Arrows indicate some remarkable fluxes (see text for details).}
\label{fig:pappbc}
\end{figure*} 
%
%

%
%
\section{Rauzy and random tilings}
%
%
Rauzy tilings can be seen as generalizations of the Fibonacci chain to higher dimensions \cite{Vidal01}. These codimension-1 quasicrystalline tilings are built using the cut-and-project method \cite{Kalugin85, Duneau85, Elser85}. In the following, we consider the two-dimensional Rauzy tiling and its approximants that are based on the Tribonacci sequence, 
%
%
\be 
R_{k+1}=R_k+R_{k-1}+R_{k-2}, \: \forall k>1 \in \mathbb{N},
\ee 
%
%
with \mbox{$R_0=R_1=1$}, and $R_2=2$. The order-$k$ approximant contains $R_{k+1}$ sites and, after a proper ordering of the sites (according to their position in the perpendicular space in the cut-and-project construction), its connectivity matrix has a Toeplitz-like structure with bands starting at positions ($R_{k-2}, R_{k-1}, R_{k}$)~\cite{Vidal01}. This rhombus tiling contains three, four, and five-fold coordinated sites. In the quasiperiodic limit, their densities are given  by \mbox{$\rho_3=\rho_5=2\,\theta^{-3}$} and \mbox{$\rho_4=1-\rho_3-\rho_5$}, where $\theta=\displaystyle{\lim_{k\rightarrow+ \infty}} R_{k+1}/R_{k} \simeq 1.839$ is the so-called \mbox{Tribonacci} constant defined as the Pisot root of the equation \mbox{$x^3=x^2+x+1$}. As any codimension-1 tiling, Rauzy tiling approximants only possess an inversion symmetry associated with the center of the one-dimensional acceptance zone (see Fig.~\ref{fig:tiling}). 

In its original construction \cite{Vidal01}, the Rauzy tiling has three different types of tiles (corresponding to the projections of the cubic-lattice faces onto the parallel space) with incommensurate areas. However, one can change the projection direction in the cut-and-project algorithm in order to obtain identical areas.  This isometric version of the Rauzy tiling displayed in Fig.~\ref{fig:tiling} (left) is especially well suited to the problem under study~\cite{Vidal04} (see below).  Moreover, we will also pay attention to a structural disorder induced by phason flips which consist in locally changing neighbors of three-fold coordinated sites as depicted in Fig.~\ref{fig:tiling} (middle).  As argued in Ref.~\cite{Destainville02} one needs to perform about $N^2/2$ random flips to fully disorder a tiling with $N$ sites. After such a rearrangement of links, one obtains a random tiling with three, four, five, and six-fold coordinated sites [see Fig.~\ref{fig:tiling} (right)].

%
%
\section{Model and symmetries}
%
%
For simplicity, we consider a single-orbital tight-binding Hamiltonian 
%
%
\be 
H=-\sum_{\langle i,j \rangle} t_{ij} \: |i\rangle \langle j|
,
\label{eq:ham}
\ee
%
%
where the sum is performed over nearest-neighbor sites. When a magnetic field $\boldsymbol{B}$ perpendicular to the tiling is introduced, the hopping term from site $i$ to site $j$ is modified according to the Peierls substitution \mbox{$t_{ij} \rightarrow t_{ij} \, {\rm e}^{-\frac{2 {\rm i} \pi}{\phi_0} \int_{\boldsymbol{r}_i}^{\boldsymbol{r}_j} d\boldsymbol{r}\cdot \boldsymbol{A}(\boldsymbol{r})}$} where $\boldsymbol{A}$ is a vector potential such that $\boldsymbol{B}=\boldsymbol{\nabla}\times \boldsymbol{A}$. In the following, we set \mbox{$t_{ij}=\hbar=1$}, $e=-2\pi$ so that the flux quantum \mbox{$\phi_0=h/|e|=1$} and the nearest-neighbor distance \mbox{$a=1$}. We also introduce the reduced flux per  plaquette \mbox{$f=\phi / \phi_0=\pm |\boldsymbol{B}| {\mathcal A}$}, where ${\mathcal A}=\frac{\sqrt{3}}{2} a^2$ is the elementary rhombus area.

Since ${\mathcal A}$ is the same for all rhombi, the spectrum of $H$ is periodic with $f$ (at least for open boundary conditions). Consequently, we can restrict our study to $f \in [0,1]$. The spectrum is also obviously unchanged when the field direction is reversed  ($f \leftrightarrow -f$). In addition, since the lattice is bipartite, the spectrum is  symmetric with respect to $0$. For periodic boundary conditions, this symmetry is broken due to odd-length paths encircling the torus that destroy bipartiteness. Similarly, in the presence of a magnetic field, fluxes are present in the torus and destroy the periodicity with $f$.  However, these two symmetry-breaking effects become negligible in the thermodynamic limit.

%
%
\section{Boundary conditions and gauge choice}
%
%
Since the pioneering work of Hofstadter on the square lattice \cite{Hofstadter76}, the spectrum of $H$ as a function of $f$, dubbed ``Hofstadter butterfly" (see Ref.~\cite{Satija16} for a recent review), has been analyzed for many periodic two-dimensional lattices (triangular \cite{Claro79}, honeycomb \cite{Rammal85}, flat-band  \cite{Aoki96}, dice \cite{Vidal98}, kagome \cite{Xiao03},...) unveiling very rich features. The simplicity of these structures allows one to study the butterfly directly in the thermodynamic limit using suitable choices for the vector \mbox{potential $\boldsymbol{A}$}. 

For quasiperiodic systems, one needs to consider a finite-size system and, for any gauge choice, two problems arise. First, the incommensurability of tile areas breaks the periodicity of the butterfly with $f$ as originally discussed in the Penrose lattice \cite{Hatakeyama89}. Second, as for any system, if one considers open boundary conditions, edge states prevent one from identifying bulk gaps properly as discussed in Refs.~\cite{Vidal04,Tran15}. 

Here, we solve these two issues by: (i) deforming the tiling to deal with identical tile areas (see discussion above) and (ii) by considering periodic boundary conditions. This latter condition induces restrictions on the accessible reduced flux $f$ (see Appendix \ref{mt} for details). Indeed, the total flux through the system must be an integer \cite{Zak64b,LL9}.  In the following, we consider a single unit cell of an approximant with periodic boundary conditions.
In the order-$k$ approximant of the isometric Rauzy tiling (or its disordered version): (i) all plaquettes have the same area, and (ii) the number of plaquettes equals the number of sites, so that $f$ has to be chosen as a multiple \mbox{of $1/R_{k+1}$}.  
Results presented below have been obtained by numerical diagonalizations of $H$.

%
%
\begin{figure}[t]
\centering
\includegraphics[width=\columnwidth]{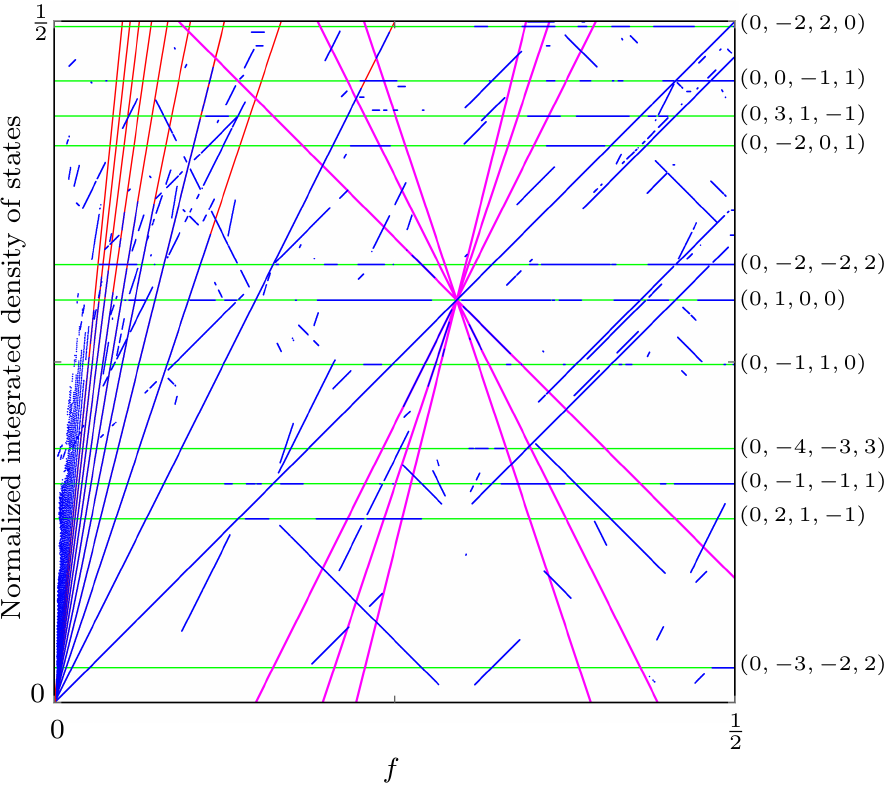}
\caption{(Color online) Wannier diagram of the isometric Rauzy tiling on a torus with $R_{16}=10609$ sites. Symmetries restrict the relevant range of  $f$ and $\mathcal{N}$ to $[0,1/2]$. Only gaps larger than $2.10^{-2}$ are shown. Gaps are labeled by four integers $(\nu,u,v,w)$. Red lines highlight the main IQHE gaps $(\nu,0,0,0)$,  magenta lines illustrate gaps $(\nu,1-\nu,0,0)$, and green lines indicate some \mbox{$(0,u,v,w)$ gaps}. Blue dots are data from diagonalizations and other lines are simply guides for the eye. This diagram and the gap labeling are unchanged for larger approximants.}
\label{fig:wannier}
\end{figure} 
%
%

%
%
\section{Hofstadter butterfly}
%
%
The zero-field energy spectrum of the isometric Rauzy tiling has been discussed in Refs.~\cite{Jagannathan01,Triozon02}. 
In the presence of a magnetic field, the spectrum has also been computed but only for open boundary conditions \cite{Vidal04,Tran15}.  In this case, edge states fill bulk gaps emerging for nonvanishing fields so that it is impossible to analyze the nontrivial characteristics of the butterfly. As shown in Fig.~\ref{fig:pappbc} (left), a very rich gap structure is unveiled when considering the system on a torus. We emphasize that all gaps visible are stable when increasing the order of the approximant so that, up to the image resolution, this butterfly should be considered as the one of the (infinite) quasiperiodic isometric Rauzy tiling.  As in most Hofstadter butterflies, one observes the presence of Landau levels arising from band edges separated by the IQHE gaps. As usual, these levels are broadened when the system is disordered [see Fig.~\ref{fig:pappbc} (right)]. Nevertheless, phason-flip disorder is sufficiently weak to preserve  main IQHE gaps while destroying the fine structure. 

%
%
\section{ Wannier diagram and gap labeling}
%
%
To proceed further, we compute the so-called Wannier diagram~\cite{Wannier78} obtained by plotting, for an energy $E$ inside a gap, the normalized integrated density of states $\mathcal{N}(E,f)$, i.e., the number of levels below $E$ divided by the total number of levels, as a function of $f$ (see Fig.~\ref{fig:wannier}). 
In the quasiperiodic limit, we conjecture that any gap can be labeled with four integers $(\nu, u, v, w)$ according to
%
%
\beq
\mathcal{N}(E,f)=\nu \, f + u\, \theta^{-2} + v \, \theta^{-1} + w.
\label{eq:conj}
\eeq 
%
 
Indeed, integrating the Widom-St$\check{\rm{r}}$eda formula~\cite{Streda82,Widom82} for the Hall conductivity at energy $E$ inside a gap,
%
%
\beq
\sigma_{\rm H} =  e  \frac{\partial \mathcal{N}(E,f)}{\partial  f}= -\frac{e^2}{h} \nu,
\eeq 
%
%
one finds $\mathcal{N}(E,f)=\nu f + \mathcal{N}_0$, where $\nu$ is a topologically-invariant integer \cite{Thouless82,Niu85} and $ \mathcal{N}_0$ is a constant. This linear dependence with $f$ is directly observed in Fig.~\ref{fig:wannier}. Note that, for open boundary conditions, $\nu$ counts the number of edge states as recently discussed in Ref.~\cite{Tran15} for the isometric Rauzy tiling. Since (i) the normalized integrated density of states is a multiple of $1/R_{k+1}$ for the order-$k$ approximant, and (ii) three consecutive Tribonacci numbers are coprime integers, B\'ezout's identity guarantees that there exists a triplet of integers $(u_k,v_k,w_k)$ such that
%
%
\beq
 \mathcal{N}_0=u_k\frac{R_{k-1}}{R_{k+1}}+v_k\frac{R_{k}}{R_{k+1}}+w_k.
\eeq
%
%
It turns out that for all the gaps we studied, we found that this triplet does not depend on $k$ so that, in the quasiperiodic limit, one gets Eq.~(\ref{eq:conj}).
This result could certainly be derived rigorously using the gap-labeling \mbox{theorem \cite{Baake_book00}}. 

Using this labeling, one can classify gaps in the Hofstadter butterfly according to $\nu$ that gives the magnetic-field dependence and to $(u,v)$ that indicate their relationship with quasiperiodicity. Indeed, as can be seen in Eq.~(\ref{eq:conj}) $u$ and $v$ are the only integers related to $\theta$ that keep track of the quasiperiodic order. The integer $w$ simply ensures that $\mathcal{N} (E,f) \in [0,1]$. Thus, one may, {\it a priori}, consider three categories: \mbox{(a) $\nu \neq 0$}, $(u,v)=(0,0)$; \mbox{(b) $\nu \neq 0$, $(u,v) \neq (0,0)$};  \mbox{(c) $\nu = 0$, $(u,v) \neq (0,0)$}. We do not consider the trivial case $\nu=u=v=0$ that corresponds to $\mathcal{N}(E,f)=0$ or $1$, i.e., a completely empty or a completely filled system.

Category (a) concerns the main IQHE gaps  (see red lines in Fig.~\ref{fig:wannier}). These gaps are robust against disorder~\cite{Thouless82} as can be seen in Fig.~\ref{fig:pappbc} and, as such, should be considered as independent of quasiperiodic order. All gaps recently identified in Ref.~\cite{Tran15} belong to this family. 

By contrast, gaps from categories (b) and (c) (green and magenta lines in Fig.~\ref{fig:wannier}) are destroyed by disorder (see Fig.~\ref{fig:pappbc}) and genuinely associated with quasiperiodic order. Gaps belonging to category (b) originate as fans, separated by Landau levels, in the vicinity of some fluxes that play a role similar to rational fluxes in the Hofstadter butterflies of periodic systems~\cite{Satija16}. In the quasiperiodic limit, these fluxes can be indexed by three integers as $p \: \theta^{-2}+q \: \theta^{-1}+ r$ [see arrows in Fig.~\ref{fig:pappbc} (left) for examples]. We note that these fluxes are also local minima of the ground-state energy as a function of $f$ as for rational fluxes in periodic systems (see Ref.~\cite{Pannetier84} for an experimental observation of this phenomenon in the square lattice).

Finally, we emphasize that, for $f=1/2$, time-reversal symmetry implies $\nu=0$ for all gaps. Remarkably, these gaps that form category (c) are also found for many other values of $f$, which is rather unusual.  Indeed, to our knowledge,  such gaps have only been observed in the Lieb \cite{Goldman11} and in the dice lattices \cite{Vidal98}. Note that, in a bipartite tight-binding model, the number of states below a gap at zero energy does not depend on $f$ ($\mathcal{N}=1/2$ for all $f$) so that $\nu=0$.

%
\section{Landau levels and effective mass} 
%
%
To better characterize the main IQHE gaps, let us focus on Landau levels that arise from band edges (see Fig.~\ref{fig:pappbc}). In the zero-flux limit, the excitation energy of the $n$th Landau level $\Delta E_n$ can be well fitted by 
%
%
\be
\Delta E_n=\hbar \frac{|e {\bf B}|}{m}\left(n+\frac{1}{2}\right)=\frac{4\pi f}{\sqrt{3} \: m}\left(n+\frac{1}{2}\right), \:  \forall n\in \mathbb{N},
\label{eq:LL}
\ee
%
%
where $1/m= 1.957(2)$ is the inverse effective mass of the electron. As expected, for the order-$k$ approximant and a given flux $f$, the degeneracy of each Landau level is given by $R_{k+1} \times f$. However, when $f$ increases, the degeneracy of these levels is lifted since lattice effects lead to a broadening as discussed in Ref.~\cite{Zak64} for crystals. 

There are several ways to understand the surprising emergence of an effective mass in nonperiodic systems (see Appendix \ref{ebem} for discussions). One possibility is to consider an infinite approximant structure with $R_{k+1}$ sites per unit cell and to compute, for $f=0$, the inverse effective mass tensor $\alpha$ of the lowest-energy band. Practically, one diagonalizes the Bloch Hamiltonian $H(\boldsymbol{k})={\rm e}^{-i\boldsymbol{k}\cdot \hat{\boldsymbol{r}}} H {\rm e}^{i\boldsymbol{k}\cdot \hat{\boldsymbol{r}}}$, where $\boldsymbol{k}$ is a Bloch wave vector and $\hat{\boldsymbol{r}}$ is the position operator. One then expands the dispersion relation of the lowest-energy band in the vicinity of $\boldsymbol{k}=(0,0)$, 
\be
\epsilon (\boldsymbol{k})\simeq \epsilon(0)+\frac{1}{2}\alpha_{ij}k_i k_j,
\ee
%
%
%
%
%
where $\epsilon(0)=-4.115008(1)$ is the ground-state energy and $\alpha$ is the inverse effective mass tensor. 
Denoting $\alpha_1$ and $\alpha_2$ its eigenvalues, the average inverse effective mass is then given by \mbox{$1/{m_\text{T}}=\sqrt{\alpha_1 \alpha_2}=1.95735(1)$}. Up to a numerical factor, the latter is equal to the dimensionless Thouless conductance \cite{Edwards72,Abrahams79} for the lowest-energy band.  Actually, it may appear fortuitous that ${m_\text{T}}$ matches $m$ so well as it is computed from the curvature of a single band whereas Landau levels are built from $R_{k+1} \times f$ bands. This result is due to a finite stiffness of the ground-state energy with respect to boundary conditions (Thouless energy \cite{Edwards72}). This nonvanishing stiffness stems from the extended nature of the ground state that we have checked explicitly. 

In the disordered case, the situation is different. As already mentioned, Landau levels broaden so that, even in the zero-flux limit, a precise determination of the effective mass is harder. On the one hand, we get $1/m\simeq 2$ by a brute-force fit of the Landau-level slope. On the other hand, for $f=0$ and since the system is disordered, one expects all states to be localized \cite{Abrahams79} and we indeed find that $1/{m_\text{T}} \propto {\rm e}^{-L/\xi}$  when increasing the linear system size $L$  with a localization length $\xi \sim 30 \: a $. A localization length much larger than the nearest-neighbor distance indicates that phason disorder should be considered as a weak disorder. When the cyclotron radius is smaller than the localization length, i.e., $f \ll \xi^{-2}$, energy levels are insensitive to the magnetic field. In the opposite case, broad Landau levels show up [see Fig.~\ref{fig:pappbc} (right)].

%
\section{Conclusion and perspectives}
%
In the present paper, we computed Hofstadter butterflies for the quasiperiodic and the disordered Rauzy tiling. In the quasiperiodic case, the butterfly displays a rich structure with three different types of gaps that can be labeled by four integers (instead of two for periodic systems as shown in Appendix \ref{app:labeling}). More generally, the number of integers needed to label gaps depends on the irrational number used in the construction of the quasicrystal. This study provides an example of a gap labeling involving both IQHE and quasiperiodicity. Large gaps seem to be associated with small integers. In the random tiling, for which quasiperiodic order is destroyed, one has $u=v=0$, and the usual labeling of IQHE gaps in terms of two integers $(\nu,w)$ is recovered.
 
For a periodic system, the Hofstadter butterfly is known to be self-similar and made of a finite (infinite) number of bands for rational (irrational) fluxes.  These properties stem from commensurability effects between the geometric and the magnetic cells that are clearly absent for a quasicrystal. However, we found that some irrational fluxes seem to play an important role. This is likely due to the self-similar property of the tiling itself but, at this stage, a complete understanding of the interplay between quasiperiodicity and magnetic field is still lacking. In particular, the nature of the spectrum as a function of the field is certainly a topic of interest. We hope that the present paper will stimulate further studies of other two-dimensional quasiperiodic systems to shed light on these issues.


\acknowledgments

We are indebted to B. Dou\c{c}ot for discussions about magnetic translations and to J. Kellendonk and R. Mosseri for insight about the gap labeling. We also thank M. Ullmo for her contribution at an early stage of this paper and acknowledge E. Akkermans, A. Jagannathan, P. Kalugin, N. Mac\'e, G. Montambaux, F. Pi\'echon, and  A. Soret for various exchanges.

%
\appendix
%
%
\section{Periodic boundary conditions and magnetic field: magnetic translations}
\label{mt}
A difficulty in computing the energy spectrum of a quasiperiodic (or random) tiling in the presence of a perpendicular magnetic field is that one has to work with a finite-size system. Indeed, one cannot use Bloch's theorem as in the case of a periodic lattice, in order to directly work in the thermodynamic limit. Working with a finite-size system, one has to make a choice for boundary conditions. Open boundary conditions are useful in the sense that any magnetic field is possible. But one drawback is that bulk levels are mixed with edge levels. In the present paper, we are interested in bulk properties and, in particular, we want to clearly identify bulk gaps. We therefore need to impose periodic boundary conditions.

A standard approach is to make a gauge choice for the vector potential (such as Landau's gauge) and then to try to impose that the Peierls phase matches the periodic boundary conditions. This is actually very inefficient and  it usually only provides a small set of allowed values of magnetic fluxes \cite{Tran15}. 

However, there is a general result  based on magnetic translations that can help us \cite{Zak64,LL9}.  Magnetic-translation operators are generalizations of the usual translation operators when a magnetic field is present. Indeed, when the hopping amplitudes of a tight-binding Hamiltonian are dressed with Peierls phases, the resulting Hamiltonian $H$ no longer commutes with translation operators $T_{\boldsymbol{a}_j}$ where $\boldsymbol{a}_1$ and $\boldsymbol{a}_2$ are two vectors defining the open boundary system (the total area of the sample being $|\boldsymbol{a}_1\times \boldsymbol{a}_2|$). This is due to  the vector potential $\boldsymbol{A}$ which is non uniform although the magnetic field is homogeneous. However, $H$ still commutes with magnetic-translation operators defined as
\be
\mathcal{T}_{\boldsymbol{a}_j}={\rm e}^{2{\rm i} \pi \chi_{\boldsymbol{a}_j}(\hat{\boldsymbol{r}})}T_{\boldsymbol{a}_j},
\ee
with 
\be
\chi_{\boldsymbol{a}_j}(\boldsymbol{r})=\int^{\boldsymbol{r}}_{\boldsymbol{0}} {\rm d}\boldsymbol{r}' \cdot [\boldsymbol{A}(\boldsymbol{r}'-\boldsymbol{a}_j)-\boldsymbol{A}(\boldsymbol{r}')],
\ee
which are the product of a gauge transformation ${\rm e}^{2{\rm i}\pi \chi_{\boldsymbol{a}_j} (\hat{\boldsymbol{r}})}$ and of a translation operator $T_{\boldsymbol{a}_j}$, where $\hat{\boldsymbol{r}}$ denotes the position operator with respect to a given origin $\boldsymbol{0}$. However, magnetic-translation operators along the two directions $\boldsymbol{a}_1$ and $\boldsymbol{a}_2$ do not commute in general. They only commute if the total magnetic flux across the sample is a multiple of the flux quantum $\phi_0=1$. For simplicity, we assume in the following that all tiles have the same area.

\begin{figure}[t]
\includegraphics[width=0.8\columnwidth]{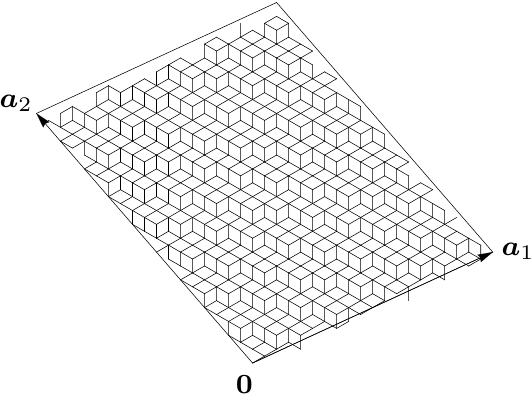}
\caption{Unit cell of the $k=10$ approximant with $R_{11}=504$ sites.}
\label{fig:ruc}
\end{figure}
%
To implement this ``magnetic-translation trick'', one thus has to proceed in several steps. First, one chooses a vector potential $\boldsymbol{A}$ and an origin $\boldsymbol{0}$. Second, one defines the two vectors $\boldsymbol{a}_1$ and $\boldsymbol{a}_2$ from the open boundary system (see Fig.~\ref{fig:ruc}). Then,   the hopping term between site $i$ and site $j$ is given by
\be
t_{ij}= {\rm e}^{2{\rm i} \pi \int_{\boldsymbol{r}_i+\boldsymbol{\tau}_i}^{\boldsymbol{r}_j+\boldsymbol{\tau}_j} {\rm d} \boldsymbol{r} \cdot \boldsymbol{A}(\boldsymbol{r}) }\:
{\rm e}^{2{\rm i} \pi \:\gamma_{i,j}},
\ee
with:
\beqn
\gamma_{ij}&=& \chi_{\varepsilon_{j,1}\boldsymbol{a}_1}({\boldsymbol{r}_j})+\chi_{\varepsilon_{j,2}\boldsymbol{a}_2}({\boldsymbol{r}_j}+\varepsilon_{j,1}\boldsymbol{a}_1) - (j\rightarrow i),\qquad \\
\boldsymbol{\tau}_i&=&\varepsilon_{i,1} \boldsymbol{a}_1+\varepsilon_{i,2} \boldsymbol{a}_2,\qquad \\
\boldsymbol{\tau}_j&=&\varepsilon_{j,1} \boldsymbol{a}_1+\varepsilon_{j,2} \boldsymbol{a}_2.\qquad
\eeqn
By convention, translation vectors $\boldsymbol{\tau}_k$ are defined  with $\varepsilon_{k,l}=0$ or 1 and obey \mbox{$|(\boldsymbol{r}_j+\boldsymbol{\tau}_j)-(\boldsymbol{r}_i+\boldsymbol{\tau}_i)|=a=1$}. Note  that to fulfill this latter condition, $\boldsymbol{\tau}_i$ and $\boldsymbol{\tau}_j$ vectors cannot be nonzero simultaneously. 
When $\boldsymbol{\tau}_i=\boldsymbol{\tau}_j=\boldsymbol{0}$, $\gamma_{i,j}$ vanishes and one recovers the usual Peierls phase. 

A possible check of the procedure consists of computing the trace of $H^4$, which counts the number of closed paths of length 4 in the tiling. When the total flux $N\times f$ is an integer, one must find:
\be
\text{Tr}\left(H^4 \right) = 8 N \cos(2\pi f)  + \text{cst},
\ee
where $N$ is the total number of plaquettes. Indeed,  there are two oriented closed paths encircling  each plaquette and each of these paths can start from any of the four vertices belonging to this plaquette, hence the factor $8 N$. The constant term simply counts the number of self-retracing paths and, as such, does not bring any dependence with $f$.

We emphasize that, contrary to the claim made in Ref.~\cite{Werner15}, this trick can be implemented for any gauge choice. However, for a finite-size system with periodic boundary conditions, different choices lead to the same magnetic flux in each plaquette but may give different fluxes through noncontractible loops of the torus. Nevertheless, closed paths associated with these loops become larger and larger when the system size increases, and the difference in the spectrum induced by these contributions vanishes in the thermodynamic limit.

\section{Zero-field  density of states and effective band edge mass}
\label{ebem}

\begin{figure}[t]
\begin{center}
\includegraphics[width=\columnwidth]{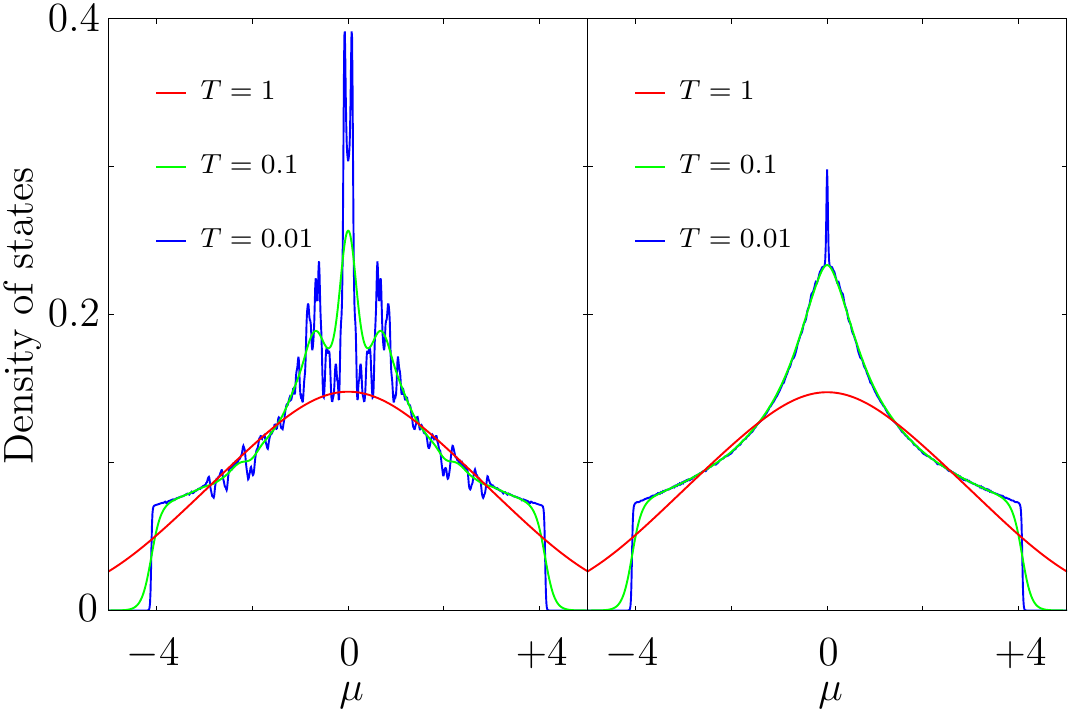}
\end{center}
\caption{Thermodynamic density of states per site $\rho(\mu,T)/N$ at temperature $T$ as a function of the chemical potential $\mu$ for  (a) the Rauzy tiling; (b) the random tiling.  Results are shown here for a system with $R_{21}=223317$ sites and should be considered as converged with the system size. The maximum relative error with the previous approximant with $R_{20}=121415$ sites being smaller than 1\% for these temperatures.}
\label{fig:dos}
\end{figure}

In this appendix, we present an alternative way to define an effective band-edge mass for a tiling. To this aim, we start by briefly discussing the zero-field thermodynamic density of states,
\be
\rho(\mu,T)=\sum_\alpha \frac{1}{4T}\text{sech}^2\frac{E_\alpha-\mu}{2T},
\ee 
plotted in Fig.~\ref{fig:dos}, where $T$ is the temperature, $\mu$ is the chemical potential controlling the electronic filling, and $\{E_\alpha\}$ are the energy eigenvalues. Here, we set the Boltzmann constant $k_{\rm B}=1$. Temperature is used to smoothen the density of states and corresponds to a box width of $\Delta E \simeq 3.53 \: T$. In the zero-temperature limit, $\rho(\mu,0)=\sum_\alpha \delta(E_\alpha-\mu)$.

At high temperatures, quasiperiodic and disordered cases coincide for all chemical potentials (see Fig.~\ref{fig:dos}). At low temperatures, the density of states remains similar (and  smooth) near the band edges, but they strongly differ near the band center. In the disordered case, the density of states is smooth,  apart from a zero-energy $\delta$ peak corresponding to less than 1\% of very localized states around six-fold coordinated sites (see Ref.~\cite{Sutherland86} for a description of these states in the dice lattice). By contrast, in the quasiperiodic case, the low-temperature density of states displays many pseudo gaps (see  Refs.~\cite{Jagannathan01,Triozon02} for the zero-temperature case).
%

In the main text, we have defined an effective (band-edge) mass from the band structure of an infinite periodic approximant. An alternative way consists of fitting the smoothed normalized integrated density of states in zero-field $\mathcal{N}(E,f=0)$ near the band edge, assuming a parabolic band edge. This amounts to write 
%
%
\be
\mathcal{N}(E,f=0)=\frac{m_\rho}{2\pi}\frac{\sqrt{3}}{2} [E-\epsilon(0)],
\label{eq:1}
\ee
%
%
where the zero-field ground-state energy  in the quasiperiodic case is \mbox{$\epsilon(0)=-4.115008(1)$} whereas $\epsilon(0)=-4.08(1)$ in the disordered case. Fitting the effective mass $m_\rho$ with this expression, one gets $1/{m_\rho}= 1.95(1)$ in the quasiperiodic case and $1/{m_\rho}= 1.9(1)$ in the disordered case. However, $m_\rho$ and $m$ discussed in the main text are identified through the Onsager semiclassical quantization of closed cyclotron orbits~\cite{Onsager52},
\be
\mathcal{N}(E_n,f=0)=\left(n+\frac{1}{2}\right)f.
\label{eq:2}
\ee 
Using Eqs.~(\ref{eq:1}) and (\ref{eq:2}), one indeed finds
\be
E_n-\epsilon(0)=\frac{4\pi f}{\sqrt{3} \: m_\rho}\left(n+\frac{1}{2}\right),  \: \forall n\in \mathbb{N},
\ee
which is similar to Eq.~(\ref{eq:LL}) provided $m_\rho=m$. We thus have two independent ways of computing this effective mass: a direct fit of the Landau levels when $f\to 0$ that gives $m$, and a fit of $\mathcal{N}(E,f=0)$ according to Eq.~(\ref{eq:1}) that gives $m_\rho$. If both approaches are in good agreement for the quasiperiodic case, results for the disordered case are less precise. A better analysis would require an average over a large number of disorder configurations but this is beyond the scope of the present paper. 

A third approach to compute an effective mass is discussed in the main text. It relies on a quadratic expansion of the lowest-energy band near its minimum and gives an inverse effective mass $1/m_{\rm T}$ equal (up to a numerical factor) to the dimensionless Thouless conductance \cite{Edwards72,Abrahams79}. For the isometric Rauzy tiling, the effective mass tensor $\alpha$ has two different eigenvalues  $\alpha_1=2.38173(1)$ and $\alpha_2=1.60857(1)$ yielding \mbox{$1/m_{\rm T}=\sqrt{\alpha_1 \alpha_2}=1.95735(1)$} and an anisotropy \mbox{$\alpha_1/ \alpha_2 =1.48065(1)$} (see Appendix \ref{app:anisotropy} for a heuristic argument).

In a periodic system, these various definitions for band-edge masses are equivalent, namely, $m=m_\rho=m_\text{T}$.  Our results suggest that it is also the case for the Rauzy tiling (up to numerical accuracy). However, this equivalence clearly no longer holds for the disordered case for which ${m}\simeq {m_\rho} \simeq 2$ but, as explained in the main text, $m_\text{T}$ diverges in the thermodynamic limit.

\section{Anisotropy of the effective mass tensor}
\label{app:anisotropy}

The inverse effective mass tensor of the lowest-energy band for the isometric Rauzy tiling features an anisotropy $\alpha_1\neq \alpha_2$ (see the previous appendix and the main text). This anisotropy stems from unequal distributions of link orientations  as we will now discuss. 

Let $|\psi\rangle$ be the ground state and let us define
\be
\tilde{\epsilon} (\boldsymbol{k})=\langle \psi |H(\boldsymbol{k})|\psi\rangle,
\ee
as an effective low-energy and long-wavelength dispersion relation (a sort of continuum limit). Note that this dispersion relation is different from the lowest-energy band defined in the main text. In particular, we expect it to approximately describe a  broader energy range above the ground state than just the lowest-energy band. In the long-wavelength limit, the above dispersion relation can be written as
\be
\tilde{\epsilon}(\boldsymbol{k})\simeq \tilde{\epsilon}(0)+\frac{1}{2}\tilde{\alpha}_{ij}k_i k_j
\ee
where $\tilde{\epsilon}(0)$ is the ground-state energy and $\tilde{\alpha}$ is an inverse effective mass tensor. 

From the exact numerical ground state (extrapolated in the thermodynamic limit), we find \mbox{$\tilde{\epsilon}(0)=\epsilon(0)=-4.11501(1)$} and an inverse effective mass tensor $\tilde{\alpha}$ with eigenvalues $\tilde{\alpha}_1=2.45515(1)$ and $\tilde{\alpha}_2=1.65985(1)$, corresponding to an average inverse effective mass $\sqrt{\tilde{\alpha}_1 \tilde{\alpha}_2}=2.01871(1)$ and an anisotropy $\tilde{\alpha}_1/\tilde{\alpha}_2=1.47914(1)$. Eigenvalues $(\tilde{\alpha}_1,\tilde{\alpha}_2)$ can be seen as approximations to $({\alpha}_1, {\alpha}_2)$ obtained by neglecting the interband contribution.

In order to gain some analytical understanding of the anisotropy and since the ground state is an extended state, we further approximate \mbox{$| \psi \rangle$} by the flat state \mbox{$|\psi\rangle \simeq \frac{1}{\sqrt{R_{k+1}}}\sum_i |i\rangle$} (for the order-$k$ approximant) and obtain
\be
\tilde{\epsilon}(\boldsymbol{k})=-\frac{1}{R_{k+1}}\sum_{\langle i,j\rangle} e^{i\boldsymbol{k}\cdot (\boldsymbol{r}_j-\boldsymbol{r}_i)}= -2\sum_{l=1}^3 t_l \cos(\boldsymbol{k} \cdot \boldsymbol{\delta}_l),
\ee
where the hopping terms $t_l$ are given by twice the density of links pointing in directions $\boldsymbol{\delta}_1,\boldsymbol{\delta}_2,\boldsymbol{\delta}_3$, namely, $(t_1,t_2,t_3)=(1-\theta^{-1},1-\theta^{-2},1-\theta^{-3})$ in the thermodynamic limit. This dispersion relation is identical to that of an anisotropic triangular lattice with hopping amplitudes $t_l$ along the three directions. In the long-wavelength limit $|\boldsymbol{k}|\ll 1/\sqrt{R_{k+1}}$, it can be approximated by a parabola, 
\be
\tilde{\epsilon}(\boldsymbol{k})\simeq \sum_{l=1}^3 t_l [-2+(\boldsymbol{k}\cdot \boldsymbol{\delta}_l)^2],
\ee
yielding  $\tilde{\epsilon}(0)=-4$ and an inverse effective mass tensor $\tilde{\alpha}$ with eigenvalues,
\be
\tilde{\alpha}_1=2+\sqrt{1-3\theta^{-2}} \text{ and } \tilde{\alpha}_2=2-\sqrt{1-3\theta^{-2}}.
\ee 
They correspond to an average inverse effective mass $\sqrt{\tilde{\alpha}_1 \tilde{\alpha}_2}\simeq 1.97$ and an anisotropy \mbox{$\tilde{\alpha}_1/\tilde{\alpha}_2\simeq1.41$}. Both quantities are in fair agreement with those derived from the lowest-energy band described in the main text. 

\section{Gap labeling for the Hofstadter butterfly of a periodic system}
\label{app:labeling}
Consider a periodic crystal with $N_u$ unit cells and $N_s$ sites per unit cell. We are interested in the thermodynamic limit where the total number of sites $N=N_{u} N_s$ goes to infinity (at fixed $N_s$). We start again by integrating the Widom-St$\check{\rm{r}}$eda formula~\cite{Streda82,Widom82} for the Hall conductivity to obtain the number of states below a given gap $\Delta$ as
%
%
\beq
N_\Delta=\nu N_\phi + N_0,
\eeq 
%
%
where $\nu$ is an integer \cite{Thouless82,Niu85}, $N_\phi=B\mathcal{A}/\phi_0$ is the number of flux quanta in the system ($\mathcal{A}$ is the total area), and $N_0$ is an integration constant. On a torus, $N_\phi$ must be an integer \cite{Dirac31}. For $B=0$, the Hamiltonian is periodic, and  Bloch's theorem indicates that each energy band contains $N_{u}$ states. Therefore, the number of states below a gap $N_0= w\: N_{u}$, where $w$ is an integer. The normalized integrated density of states in the gap $\Delta$ is $\mathcal{N}=N_\Delta/N$ so that
%
%
\beq
\mathcal{N}=\nu \bar{f} + \frac{w}{N_s},
\eeq 
%
%
where $\bar{f}=N_\phi/N$ is the average number of flux quanta per site. Any gap is therefore labeled by only two integers $(\nu,w)$ (see Ref.~\cite{Dana85} for the case $N_s=1$). In particular, we note that this labeling works for a periodic system with tiles of commensurate or incommensurate areas.


\begin{thebibliography}{48}%
\makeatletter
\providecommand \@ifxundefined [1]{%
 \@ifx{#1\undefined}
}%
\providecommand \@ifnum [1]{%
 \ifnum #1\expandafter \@firstoftwo
 \else \expandafter \@secondoftwo
 \fi
}%
\providecommand \@ifx [1]{%
 \ifx #1\expandafter \@firstoftwo
 \else \expandafter \@secondoftwo
 \fi
}%
\providecommand \natexlab [1]{#1}%
\providecommand \enquote  [1]{``#1''}%
\providecommand \bibnamefont  [1]{#1}%
\providecommand \bibfnamefont [1]{#1}%
\providecommand \citenamefont [1]{#1}%
\providecommand \href@noop [0]{\@secondoftwo}%
\providecommand \href [0]{\begingroup \@sanitize@url \@href}%
\providecommand \@href[1]{\@@startlink{#1}\@@href}%
\providecommand \@@href[1]{\endgroup#1\@@endlink}%
\providecommand \@sanitize@url [0]{\catcode `\\12\catcode `\$12\catcode
  `\&12\catcode `\#12\catcode `\^12\catcode `\_12\catcode `\%12\relax}%
\providecommand \@@startlink[1]{}%
\providecommand \@@endlink[0]{}%
\providecommand \url  [0]{\begingroup\@sanitize@url \@url }%
\providecommand \@url [1]{\endgroup\@href {#1}{\urlprefix }}%
\providecommand \urlprefix  [0]{URL }%
\providecommand \Eprint [0]{\href }%
\providecommand \doibase [0]{http://dx.doi.org/}%
\providecommand \selectlanguage [0]{\@gobble}%
\providecommand \bibinfo  [0]{\@secondoftwo}%
\providecommand \bibfield  [0]{\@secondoftwo}%
\providecommand \translation [1]{[#1]}%
\providecommand \BibitemOpen [0]{}%
\providecommand \bibitemStop [0]{}%
\providecommand \bibitemNoStop [0]{.\EOS\space}%
\providecommand \EOS [0]{\spacefactor3000\relax}%
\providecommand \BibitemShut  [1]{\csname bibitem#1\endcsname}%
\let\auto@bib@innerbib\@empty
\bibitem [{\citenamefont {Shechtman}\ \emph {et~al.}(1984)\citenamefont
  {Shechtman}, \citenamefont {Blech}, \citenamefont {Gratias},\ and\
  \citenamefont {Cahn}}]{Shechtman84}%
  \BibitemOpen
  \bibfield  {author} {\bibinfo {author} {\bibfnamefont {D.}~\bibnamefont
  {Shechtman}}, \bibinfo {author} {\bibfnamefont {I.}~\bibnamefont {Blech}},
  \bibinfo {author} {\bibfnamefont {D.}~\bibnamefont {Gratias}}, \ and\
  \bibinfo {author} {\bibfnamefont {J.~W.}\ \bibnamefont {Cahn}},\ }\bibfield
  {title} {\enquote {\bibinfo {title} {{Metallic Phase with Long-Range
  Orientational Order and No Translational Symmetry}},}\ }\href {\doibase
  10.1103/PhysRevLett.53.1951} {\bibfield  {journal} {\bibinfo  {journal}
  {Phys. Rev. Lett.}\ }\textbf {\bibinfo {volume} {53}},\ \bibinfo {pages}
  {1951} (\bibinfo {year} {1984})}\BibitemShut {NoStop}%
\bibitem [{\citenamefont {Steurer}\ and\ \citenamefont
  {Sutter-Widmer}(2007)}]{Steurer07}%
  \BibitemOpen
  \bibfield  {author} {\bibinfo {author} {\bibfnamefont {W.}~\bibnamefont
  {Steurer}}\ and\ \bibinfo {author} {\bibfnamefont {D.}~\bibnamefont
  {Sutter-Widmer}},\ }\bibfield  {title} {\enquote {\bibinfo {title} {{Photonic
  and phononic quasicrystals}},}\ }\href {\doibase 10.1088/0022-3727/40/13/R01}
  {\bibfield  {journal} {\bibinfo  {journal} {J. Phys. D}\ }\textbf {\bibinfo
  {volume} {40}},\ \bibinfo {pages} {R229} (\bibinfo {year}
  {2007})}\BibitemShut {NoStop}%
\bibitem [{\citenamefont {Guidoni}\ \emph {et~al.}(1999)\citenamefont
  {Guidoni}, \citenamefont {D\'epret}, \citenamefont {di~Stefano},\ and\
  \citenamefont {Verkerk}}]{Verkerk99}%
  \BibitemOpen
  \bibfield  {author} {\bibinfo {author} {\bibfnamefont {L.}~\bibnamefont
  {Guidoni}}, \bibinfo {author} {\bibfnamefont {B.}~\bibnamefont {D\'epret}},
  \bibinfo {author} {\bibfnamefont {A.}~\bibnamefont {di~Stefano}}, \ and\
  \bibinfo {author} {\bibfnamefont {P.}~\bibnamefont {Verkerk}},\ }\bibfield
  {title} {\enquote {\bibinfo {title} {{Atomic diffusion in an optical
  quasicrystal with five-fold symmetry}},}\ }\href {\doibase
  10.1103/PhysRevA.60.R4233} {\bibfield  {journal} {\bibinfo  {journal} {Phys.
  Rev. A}\ }\textbf {\bibinfo {volume} {60}},\ \bibinfo {pages} {R4233(R)}
  (\bibinfo {year} {1999})}\BibitemShut {NoStop}%
\bibitem [{\citenamefont {Kraus}\ \emph {et~al.}(2012)\citenamefont {Kraus},
  \citenamefont {Lahini}, \citenamefont {Ringel}, \citenamefont {Verbin},\ and\
  \citenamefont {Zilberberg}}]{Krauss12}%
  \BibitemOpen
  \bibfield  {author} {\bibinfo {author} {\bibfnamefont {Y.~E.}\ \bibnamefont
  {Kraus}}, \bibinfo {author} {\bibfnamefont {Y.}~\bibnamefont {Lahini}},
  \bibinfo {author} {\bibfnamefont {Z.}~\bibnamefont {Ringel}}, \bibinfo
  {author} {\bibfnamefont {M.}~\bibnamefont {Verbin}}, \ and\ \bibinfo {author}
  {\bibfnamefont {O.}~\bibnamefont {Zilberberg}},\ }\bibfield  {title}
  {\enquote {\bibinfo {title} {{Topological States and Adiabatic Pumping in
  Quasicrystals}},}\ }\href {\doibase 10.1103/PhysRevLett.109.106402}
  {\bibfield  {journal} {\bibinfo  {journal} {Phys. Rev. Lett.}\ }\textbf
  {\bibinfo {volume} {109}},\ \bibinfo {pages} {106402} (\bibinfo {year}
  {2012})}\BibitemShut {NoStop}%
\bibitem [{\citenamefont {Bandres}\ \emph {et~al.}(2016)\citenamefont
  {Bandres}, \citenamefont {Rechtsman},\ and\ \citenamefont
  {Segev}}]{Bandres16}%
  \BibitemOpen
  \bibfield  {author} {\bibinfo {author} {\bibfnamefont {M.~A.}\ \bibnamefont
  {Bandres}}, \bibinfo {author} {\bibfnamefont {M.~C.}\ \bibnamefont
  {Rechtsman}}, \ and\ \bibinfo {author} {\bibfnamefont {M.}~\bibnamefont
  {Segev}},\ }\bibfield  {title} {\enquote {\bibinfo {title} {{Topological
  Photonic Quasicrystals: Fractal Topological Spectrum and Protected
  Transport}},}\ }\href {\doibase 10.1103/PhysRevX.6.011016} {\bibfield
  {journal} {\bibinfo  {journal} {Phys. Rev. X}\ }\textbf {\bibinfo {volume}
  {6}},\ \bibinfo {pages} {011016} (\bibinfo {year} {2016})}\BibitemShut
  {NoStop}%
\bibitem [{\citenamefont {Tanese}\ \emph {et~al.}(2014)\citenamefont {Tanese},
  \citenamefont {Gurevich}, \citenamefont {Baboux}, \citenamefont {Jacqmin},
  \citenamefont {Lema\^itre}, \citenamefont {Galopin}, \citenamefont {Sagnes},
  \citenamefont {Amo}, \citenamefont {Bloch},\ and\ \citenamefont
  {Akkermans}}]{Tanese14}%
  \BibitemOpen
  \bibfield  {author} {\bibinfo {author} {\bibfnamefont {D.}~\bibnamefont
  {Tanese}}, \bibinfo {author} {\bibfnamefont {E.}~\bibnamefont {Gurevich}},
  \bibinfo {author} {\bibfnamefont {F.}~\bibnamefont {Baboux}}, \bibinfo
  {author} {\bibfnamefont {T.}~\bibnamefont {Jacqmin}}, \bibinfo {author}
  {\bibfnamefont {A.}~\bibnamefont {Lema\^itre}}, \bibinfo {author}
  {\bibfnamefont {E.}~\bibnamefont {Galopin}}, \bibinfo {author} {\bibfnamefont
  {I.}~\bibnamefont {Sagnes}}, \bibinfo {author} {\bibfnamefont
  {A.}~\bibnamefont {Amo}}, \bibinfo {author} {\bibfnamefont {J.}~\bibnamefont
  {Bloch}}, \ and\ \bibinfo {author} {\bibfnamefont {E.}~\bibnamefont
  {Akkermans}},\ }\bibfield  {title} {\enquote {\bibinfo {title} {{Fractal
  Energy Spectrum of a Polariton Gas in a Fibonacci Quasiperiodic
  Potential}},}\ }\href {\doibase 10.1103/PhysRevLett.112.146404} {\bibfield
  {journal} {\bibinfo  {journal} {Phys. Rev. Lett.}\ }\textbf {\bibinfo
  {volume} {112}},\ \bibinfo {pages} {146404} (\bibinfo {year}
  {2014})}\BibitemShut {NoStop}%
\bibitem [{\citenamefont {Vignolo}\ \emph {et~al.}(2016)\citenamefont
  {Vignolo}, \citenamefont {Bellec}, \citenamefont {B\"ohm}, \citenamefont
  {Camara}, \citenamefont {Gambaudo}, \citenamefont {Kuhl},\ and\ \citenamefont
  {Mortessagne}}]{Vignolo16}%
  \BibitemOpen
  \bibfield  {author} {\bibinfo {author} {\bibfnamefont {P.}~\bibnamefont
  {Vignolo}}, \bibinfo {author} {\bibfnamefont {M.}~\bibnamefont {Bellec}},
  \bibinfo {author} {\bibfnamefont {J.}~\bibnamefont {B\"ohm}}, \bibinfo
  {author} {\bibfnamefont {A.}~\bibnamefont {Camara}}, \bibinfo {author}
  {\bibfnamefont {J.~M.}\ \bibnamefont {Gambaudo}}, \bibinfo {author}
  {\bibfnamefont {U.}~\bibnamefont {Kuhl}}, \ and\ \bibinfo {author}
  {\bibfnamefont {F.}~\bibnamefont {Mortessagne}},\ }\bibfield  {title}
  {\enquote {\bibinfo {title} {{Energy landscape in a Penrose tiling}},}\
  }\href {\doibase 10.1103/PhysRevB.93.075141} {\bibfield  {journal} {\bibinfo
  {journal} {Phys. Rev. B}\ }\textbf {\bibinfo {volume} {93}},\ \bibinfo
  {pages} {075141} (\bibinfo {year} {2016})}\BibitemShut {NoStop}%
\bibitem [{\citenamefont {Thouless}(1983)}]{Thouless83b}%
  \BibitemOpen
  \bibfield  {author} {\bibinfo {author} {\bibfnamefont {D.~J.}\ \bibnamefont
  {Thouless}},\ }\bibfield  {title} {\enquote {\bibinfo {title} {{Quantization
  of particle transport}},}\ }\href {\doibase 10.1103/PhysRevB.27.6083}
  {\bibfield  {journal} {\bibinfo  {journal} {Phys. Rev. B}\ }\textbf {\bibinfo
  {volume} {27}},\ \bibinfo {pages} {6083} (\bibinfo {year}
  {1983})}\BibitemShut {NoStop}%
\bibitem [{\citenamefont {Kunz}(1986)}]{Kunz86}%
  \BibitemOpen
  \bibfield  {author} {\bibinfo {author} {\bibfnamefont {H.}~\bibnamefont
  {Kunz}},\ }\bibfield  {title} {\enquote {\bibinfo {title} {{Quantized
  Currents and Topological Invariants for Electrons in Incommensurate
  Potentials}},}\ }\href {\doibase 10.1103/PhysRevLett.57.1095} {\bibfield
  {journal} {\bibinfo  {journal} {Phys. Rev. Lett.}\ }\textbf {\bibinfo
  {volume} {57}},\ \bibinfo {pages} {1095} (\bibinfo {year}
  {1986})}\BibitemShut {NoStop}%
\bibitem [{\citenamefont {Hofstadter}(1976)}]{Hofstadter76}%
  \BibitemOpen
  \bibfield  {author} {\bibinfo {author} {\bibfnamefont {D.~R.}\ \bibnamefont
  {Hofstadter}},\ }\bibfield  {title} {\enquote {\bibinfo {title} {{Energy
  levels and wave functions of Bloch electrons in rational and irrational
  magnetic fields}},}\ }\href {\doibase 10.1103/PhysRevB.14.2239} {\bibfield
  {journal} {\bibinfo  {journal} {Phys. Rev.}\ }\textbf {\bibinfo {volume}
  {14}},\ \bibinfo {pages} {2239} (\bibinfo {year} {1976})}\BibitemShut
  {NoStop}%
\bibitem [{\citenamefont {Harper}(1955)}]{Harper55}%
  \BibitemOpen
  \bibfield  {author} {\bibinfo {author} {\bibfnamefont {P.~G.}\ \bibnamefont
  {Harper}},\ }\bibfield  {title} {\enquote {\bibinfo {title} {{Single Band
  Motion of Conduction Electrons in a Uniform Magnetic Field}},}\ }\href
  {\doibase 10.1088/0370-1298/68/10/304} {\bibfield  {journal} {\bibinfo
  {journal} {Proc. Phys. Soc. A}\ }\textbf {\bibinfo {volume} {68}},\ \bibinfo
  {pages} {874} (\bibinfo {year} {1955})}\BibitemShut {NoStop}%
\bibitem [{\citenamefont {Aubry}\ and\ \citenamefont {Andr\'e}()}]{Aubry80}%
  \BibitemOpen
  \bibfield  {author} {\bibinfo {author} {\bibfnamefont {S.}~\bibnamefont
  {Aubry}}\ and\ \bibinfo {author} {\bibfnamefont {G.}~\bibnamefont
  {Andr\'e}},\ }\href@noop {} {\enquote {\bibinfo {title} {{Analyticity
  breaking and Anderson localization in incommensurate lattices}},}\ }\bibinfo
  {note}
  {\href{https://www.researchgate.net/publication/265502988_Analyticity_breaking_and_Anderson_localization_in_incommensurate_lattices}{Ann.
  Isr. Phys. Soc. {\bf 3}, 133 (1980)}}\BibitemShut {NoStop}%
\bibitem [{\citenamefont {Levy}\ \emph {et~al.}()\citenamefont {Levy},
  \citenamefont {Barak}, \citenamefont {Fisher},\ and\ \citenamefont
  {Akkermans}}]{Levy15}%
  \BibitemOpen
  \bibfield  {author} {\bibinfo {author} {\bibfnamefont {E.}~\bibnamefont
  {Levy}}, \bibinfo {author} {\bibfnamefont {A.}~\bibnamefont {Barak}},
  \bibinfo {author} {\bibfnamefont {A.}~\bibnamefont {Fisher}}, \ and\ \bibinfo
  {author} {\bibfnamefont {E.}~\bibnamefont {Akkermans}},\ }\href@noop {}
  {\enquote {\bibinfo {title} {{Topological properties of Fibonacci
  quasicrystals : A scattering analysis of Chern numbers}},}\ }\bibinfo {note}
  {\href{http://arxiv.org/abs/arXiv:1509.04028}{arXiv:1509.04028}}\BibitemShut
  {NoStop}%
\bibitem [{\citenamefont {Fulga}\ \emph {et~al.}(2016)\citenamefont {Fulga},
  \citenamefont {Pikulin},\ and\ \citenamefont {Loring}}]{Fulga16}%
  \BibitemOpen
  \bibfield  {author} {\bibinfo {author} {\bibfnamefont {I.~C.}\ \bibnamefont
  {Fulga}}, \bibinfo {author} {\bibfnamefont {D.~I.}\ \bibnamefont {Pikulin}},
  \ and\ \bibinfo {author} {\bibfnamefont {T.~A.}\ \bibnamefont {Loring}},\
  }\bibfield  {title} {\enquote {\bibinfo {title} {{Aperiodic Weak Topological
  Superconductors}},}\ }\href {\doibase 10.1103/PhysRevLett.116.257002}
  {\bibfield  {journal} {\bibinfo  {journal} {Phys. Rev. Lett.}\ }\textbf
  {\bibinfo {volume} {116}},\ \bibinfo {pages} {257002} (\bibinfo {year}
  {2016})}\BibitemShut {NoStop}%
\bibitem [{\citenamefont {Hatakeyama}\ and\ \citenamefont
  {Kamimura}(1989)}]{Hatakeyama89}%
  \BibitemOpen
  \bibfield  {author} {\bibinfo {author} {\bibfnamefont {T.}~\bibnamefont
  {Hatakeyama}}\ and\ \bibinfo {author} {\bibfnamefont {H.}~\bibnamefont
  {Kamimura}},\ }\bibfield  {title} {\enquote {\bibinfo {title} {{Fractal
  Nature of the Electronique Structure of a Penrose Tiling Lattice in a
  Magnetic Field}},}\ }\href {\doibase 10.1143/JPSJ.58.260} {\bibfield
  {journal} {\bibinfo  {journal} {J. Phys. Soc. Jpn.}\ }\textbf {\bibinfo
  {volume} {58}},\ \bibinfo {pages} {260} (\bibinfo {year} {1989})}\BibitemShut
  {NoStop}%
\bibitem [{\citenamefont {Vidal}\ and\ \citenamefont
  {Mosseri}(2004)}]{Vidal04}%
  \BibitemOpen
  \bibfield  {author} {\bibinfo {author} {\bibfnamefont {J.}~\bibnamefont
  {Vidal}}\ and\ \bibinfo {author} {\bibfnamefont {R.}~\bibnamefont
  {Mosseri}},\ }\bibfield  {title} {\enquote {\bibinfo {title} {{Quasiperiodic
  tilings under magnetic field}},}\ }\href {\doibase
  10.1016/j.jnoncrysol.2003.11.027} {\bibfield  {journal} {\bibinfo  {journal}
  {J. Non-Cryst. Solids}\ }\textbf {\bibinfo {volume} {334}},\ \bibinfo {pages}
  {130} (\bibinfo {year} {2004})}\BibitemShut {NoStop}%
\bibitem [{\citenamefont {Tran}\ \emph {et~al.}(2015)\citenamefont {Tran},
  \citenamefont {Dauphin}, \citenamefont {Goldman},\ and\ \citenamefont
  {Gaspard}}]{Tran15}%
  \BibitemOpen
  \bibfield  {author} {\bibinfo {author} {\bibfnamefont {D.-T.}\ \bibnamefont
  {Tran}}, \bibinfo {author} {\bibfnamefont {A.}~\bibnamefont {Dauphin}},
  \bibinfo {author} {\bibfnamefont {N.}~\bibnamefont {Goldman}}, \ and\
  \bibinfo {author} {\bibfnamefont {P.}~\bibnamefont {Gaspard}},\ }\bibfield
  {title} {\enquote {\bibinfo {title} {{Topological Hofstadter insulators in a
  two-dimensional quasicrystal}},}\ }\href {\doibase
  10.1103/PhysRevB.91.085125} {\bibfield  {journal} {\bibinfo  {journal} {Phys.
  Rev. B}\ }\textbf {\bibinfo {volume} {91}},\ \bibinfo {pages} {085125}
  (\bibinfo {year} {2015})}\BibitemShut {NoStop}%
\bibitem [{\citenamefont {Satija}()}]{Satija16}%
  \BibitemOpen
  \bibfield  {author} {\bibinfo {author} {\bibfnamefont {I.~I.}\ \bibnamefont
  {Satija}},\ }\href@noop {} {\enquote {\bibinfo {title} {Butterfly in the
  quantum world},}\ }\bibinfo {note}
  {\href{http://dx.doi.org/10.1088/978-1-6817-4117-8}{(Morgan \& Claypool
  Publishers, San Raphael, 2016)}}\BibitemShut {NoStop}%
\bibitem [{\citenamefont {Vidal}\ and\ \citenamefont
  {Mosseri}(2001)}]{Vidal01}%
  \BibitemOpen
  \bibfield  {author} {\bibinfo {author} {\bibfnamefont {J.}~\bibnamefont
  {Vidal}}\ and\ \bibinfo {author} {\bibfnamefont {R.}~\bibnamefont
  {Mosseri}},\ }\bibfield  {title} {\enquote {\bibinfo {title} {{Generalized
  quasiperiodic Rauzy tilings}},}\ }\href {\doibase
  10.1088/0305-4470/34/18/317} {\bibfield  {journal} {\bibinfo  {journal} {J.
  Phys. A}\ }\textbf {\bibinfo {volume} {34}},\ \bibinfo {pages} {3927}
  (\bibinfo {year} {2001})}\BibitemShut {NoStop}%
\bibitem [{\citenamefont {Kalugin}\ \emph {et~al.}(1985)\citenamefont
  {Kalugin}, \citenamefont {Kitaev},\ and\ \citenamefont
  {Levitov}}]{Kalugin85}%
  \BibitemOpen
  \bibfield  {author} {\bibinfo {author} {\bibfnamefont {P.~A.}\ \bibnamefont
  {Kalugin}}, \bibinfo {author} {\bibfnamefont {A.~Yu.}\ \bibnamefont
  {Kitaev}}, \ and\ \bibinfo {author} {\bibfnamefont {L.~S.}\ \bibnamefont
  {Levitov}},\ }\bibfield  {title} {\enquote {\bibinfo {title}
  {{Al$_{0.86}$Mn$_{0.14}$: a six-dimensional crystal}},}\ }\href {\doibase
  http://www.jetpletters.ac.ru/ps/1442/article_21941.shtml} {\bibfield
  {journal} {\bibinfo  {journal} {JETP Lett.}\ }\textbf {\bibinfo {volume}
  {41}},\ \bibinfo {pages} {145} (\bibinfo {year} {1985})}\BibitemShut
  {NoStop}%
\bibitem [{\citenamefont {Duneau}\ and\ \citenamefont {Katz}(1985)}]{Duneau85}%
  \BibitemOpen
  \bibfield  {author} {\bibinfo {author} {\bibfnamefont {M.}~\bibnamefont
  {Duneau}}\ and\ \bibinfo {author} {\bibfnamefont {A.}~\bibnamefont {Katz}},\
  }\bibfield  {title} {\enquote {\bibinfo {title} {{Quasiperiodic Patterns}},}\
  }\href {\doibase 10.1103/PhysRevLett.54.2688} {\bibfield  {journal} {\bibinfo
   {journal} {Phys. Rev. Lett.}\ }\textbf {\bibinfo {volume} {54}},\ \bibinfo
  {pages} {2688} (\bibinfo {year} {1985})}\BibitemShut {NoStop}%
\bibitem [{\citenamefont {Elser}(1986)}]{Elser85}%
  \BibitemOpen
  \bibfield  {author} {\bibinfo {author} {\bibfnamefont {V.}~\bibnamefont
  {Elser}},\ }\bibfield  {title} {\enquote {\bibinfo {title} {{The Diffraction
  Pattern of Projected Structures}},}\ }\href {\doibase
  10.1107/S0108767386099932} {\bibfield  {journal} {\bibinfo  {journal} {Acta
  Crystallogr., Sect. A: Found. Crystallogr.}\ }\textbf {\bibinfo {volume}
  {42}},\ \bibinfo {pages} {36} (\bibinfo {year} {1986})}\BibitemShut {NoStop}%
\bibitem [{\citenamefont {Destainville}(2002)}]{Destainville02}%
  \BibitemOpen
  \bibfield  {author} {\bibinfo {author} {\bibfnamefont {N.}~\bibnamefont
  {Destainville}},\ }\bibfield  {title} {\enquote {\bibinfo {title} {{Flip
  Dynamics in Octagonal Rhombus Tiling Sets}},}\ }\href {\doibase
  10.1103/PhysRevLett.88.030601} {\bibfield  {journal} {\bibinfo  {journal}
  {Phys. Rev. Lett.}\ }\textbf {\bibinfo {volume} {88}},\ \bibinfo {pages}
  {030601} (\bibinfo {year} {2002})}\BibitemShut {NoStop}%
\bibitem [{\citenamefont {Claro}\ and\ \citenamefont
  {Wannier}(1979)}]{Claro79}%
  \BibitemOpen
  \bibfield  {author} {\bibinfo {author} {\bibfnamefont {F.~H.}\ \bibnamefont
  {Claro}}\ and\ \bibinfo {author} {\bibfnamefont {G.~H.}\ \bibnamefont
  {Wannier}},\ }\bibfield  {title} {\enquote {\bibinfo {title} {Magnetic
  subband structure of electrons in hexagonal lattices},}\ }\href {\doibase
  10.1103/PhysRevB.19.6068} {\bibfield  {journal} {\bibinfo  {journal} {Phys.
  Rev. B}\ }\textbf {\bibinfo {volume} {19}},\ \bibinfo {pages} {6068}
  (\bibinfo {year} {1979})}\BibitemShut {NoStop}%
\bibitem [{\citenamefont {Rammal}(1985)}]{Rammal85}%
  \BibitemOpen
  \bibfield  {author} {\bibinfo {author} {\bibfnamefont {R.}~\bibnamefont
  {Rammal}},\ }\bibfield  {title} {\enquote {\bibinfo {title} {{Landau level
  spectrum of Bloch electrons in a honeycomb lattice}},}\ }\href {\doibase
  10.1051/jphys:019850046080134500} {\bibfield  {journal} {\bibinfo  {journal}
  {J. Phys. (Paris)}\ }\textbf {\bibinfo {volume} {46}},\ \bibinfo {pages}
  {1345} (\bibinfo {year} {1985})}\BibitemShut {NoStop}%
\bibitem [{\citenamefont {Aoki}\ \emph {et~al.}(1996)\citenamefont {Aoki},
  \citenamefont {Ando},\ and\ \citenamefont {Matsumura}}]{Aoki96}%
  \BibitemOpen
  \bibfield  {author} {\bibinfo {author} {\bibfnamefont {H.}~\bibnamefont
  {Aoki}}, \bibinfo {author} {\bibfnamefont {M.}~\bibnamefont {Ando}}, \ and\
  \bibinfo {author} {\bibfnamefont {H.}~\bibnamefont {Matsumura}},\ }\bibfield
  {title} {\enquote {\bibinfo {title} {{Hofstadter butterflies for flat
  bands}},}\ }\href {\doibase 10.1103/PhysRevB.54.R17296} {\bibfield  {journal}
  {\bibinfo  {journal} {Phys. Rev. B}\ }\textbf {\bibinfo {volume} {54}},\
  \bibinfo {pages} {R17296} (\bibinfo {year} {1996})}\BibitemShut {NoStop}%
\bibitem [{\citenamefont {Vidal}\ \emph {et~al.}(1998)\citenamefont {Vidal},
  \citenamefont {Mosseri},\ and\ \citenamefont {Dou\c{c}ot}}]{Vidal98}%
  \BibitemOpen
  \bibfield  {author} {\bibinfo {author} {\bibfnamefont {J.}~\bibnamefont
  {Vidal}}, \bibinfo {author} {\bibfnamefont {R.}~\bibnamefont {Mosseri}}, \
  and\ \bibinfo {author} {\bibfnamefont {B.}~\bibnamefont {Dou\c{c}ot}},\
  }\bibfield  {title} {\enquote {\bibinfo {title} {{Aharonov-Bohm Cages in
  Two-Dimensional Structures}},}\ }\href {\doibase 10.1103/PhysRevLett.81.5888}
  {\bibfield  {journal} {\bibinfo  {journal} {Phys. Rev. Lett.}\ }\textbf
  {\bibinfo {volume} {81}},\ \bibinfo {pages} {5888} (\bibinfo {year}
  {1998})}\BibitemShut {NoStop}%
\bibitem [{\citenamefont {Xiao}\ \emph {et~al.}(2003)\citenamefont {Xiao},
  \citenamefont {Pelletier}, \citenamefont {Chaikin},\ and\ \citenamefont
  {Huse}}]{Xiao03}%
  \BibitemOpen
  \bibfield  {author} {\bibinfo {author} {\bibfnamefont {Y.}~\bibnamefont
  {Xiao}}, \bibinfo {author} {\bibfnamefont {V.}~\bibnamefont {Pelletier}},
  \bibinfo {author} {\bibfnamefont {P.~M.}\ \bibnamefont {Chaikin}}, \ and\
  \bibinfo {author} {\bibfnamefont {D.~A.}\ \bibnamefont {Huse}},\ }\bibfield
  {title} {\enquote {\bibinfo {title} {{Landau levels in the case of two
  degenerate coupled bands: Kagom\'e lattice tight-binding spectrum}},}\ }\href
  {\doibase 10.1103/PhysRevB.67.104505} {\bibfield  {journal} {\bibinfo
  {journal} {Phys. Rev. B}\ }\textbf {\bibinfo {volume} {67}},\ \bibinfo
  {pages} {104505} (\bibinfo {year} {2003})}\BibitemShut {NoStop}%
\bibitem [{\citenamefont {Zak}(1964{\natexlab{a}})}]{Zak64b}%
  \BibitemOpen
  \bibfield  {author} {\bibinfo {author} {\bibfnamefont {J.}~\bibnamefont
  {Zak}},\ }\bibfield  {title} {\enquote {\bibinfo {title} {{Magnetic
  Translation Group}},}\ }\href {\doibase 10.1103/PhysRev.134.A1602} {\bibfield
   {journal} {\bibinfo  {journal} {Phys. Rev.}\ }\textbf {\bibinfo {volume}
  {134}},\ \bibinfo {pages} {1602} (\bibinfo {year}
  {1964}{\natexlab{a}})}\BibitemShut {NoStop}%
\bibitem [{\citenamefont {Lifshitz}\ and\ \citenamefont {Pitaevskii}()}]{LL9}%
  \BibitemOpen
  \bibfield  {author} {\bibinfo {author} {\bibfnamefont {E.~M.}\ \bibnamefont
  {Lifshitz}}\ and\ \bibinfo {author} {\bibfnamefont {L.~P.}\ \bibnamefont
  {Pitaevskii}},\ }\href@noop {} {\enquote {\bibinfo {title} {{\it Statistical
  Physics, Part 2: Theory of the Condensed State}},}\ }\bibinfo {note} {1st ed.
  (Butterworth-Heinemann, Oxford, 1980) Vol. 9, Sec. 60}\BibitemShut {NoStop}%
\bibitem [{\citenamefont {Jagannathan}(2001)}]{Jagannathan01}%
  \BibitemOpen
  \bibfield  {author} {\bibinfo {author} {\bibfnamefont {A.}~\bibnamefont
  {Jagannathan}},\ }\bibfield  {title} {\enquote {\bibinfo {title} {{Less
  singular quasicrystals: The case of low codimensions}},}\ }\href {\doibase
  10.1103/PhysRevB.64.140201} {\bibfield  {journal} {\bibinfo  {journal} {Phys.
  Rev. B}\ }\textbf {\bibinfo {volume} {64}},\ \bibinfo {pages} {140201}
  (\bibinfo {year} {2001})}\BibitemShut {NoStop}%
\bibitem [{\citenamefont {Triozon}\ \emph {et~al.}(2002)\citenamefont
  {Triozon}, \citenamefont {Vidal}, \citenamefont {Mosseri},\ and\
  \citenamefont {Mayou}}]{Triozon02}%
  \BibitemOpen
  \bibfield  {author} {\bibinfo {author} {\bibfnamefont {F.}~\bibnamefont
  {Triozon}}, \bibinfo {author} {\bibfnamefont {J.}~\bibnamefont {Vidal}},
  \bibinfo {author} {\bibfnamefont {R.}~\bibnamefont {Mosseri}}, \ and\
  \bibinfo {author} {\bibfnamefont {D.}~\bibnamefont {Mayou}},\ }\bibfield
  {title} {\enquote {\bibinfo {title} {{Quantum dynamics in two- and
  three-dimensional quasiperiodic tilings}},}\ }\href {\doibase
  10.1103/PhysRevB.65.220202} {\bibfield  {journal} {\bibinfo  {journal} {Phys.
  Rev. B}\ }\textbf {\bibinfo {volume} {65}},\ \bibinfo {pages} {220202}
  (\bibinfo {year} {2002})}\BibitemShut {NoStop}%
\bibitem [{\citenamefont {Wannier}(1978)}]{Wannier78}%
  \BibitemOpen
  \bibfield  {author} {\bibinfo {author} {\bibfnamefont {G.~H.}\ \bibnamefont
  {Wannier}},\ }\bibfield  {title} {\enquote {\bibinfo {title} {{A Result Not
  Dependent on Rationality for Bloch Electrons in a Magnetic Field}},}\ }\href
  {\doibase 10.1002/pssb.2220880243} {\bibfield  {journal} {\bibinfo  {journal}
  {Phys. Stat. Sol. (b)}\ }\textbf {\bibinfo {volume} {88}},\ \bibinfo {pages}
  {757} (\bibinfo {year} {1978})}\BibitemShut {NoStop}%
\bibitem [{\citenamefont {St$\check{\rm{r}}$eda}(1982)}]{Streda82}%
  \BibitemOpen
  \bibfield  {author} {\bibinfo {author} {\bibfnamefont {P.}~\bibnamefont
  {St$\check{\rm{r}}$eda}},\ }\bibfield  {title} {\enquote {\bibinfo {title}
  {{Quantised Hall effect in a two-dimensional periodic potential}},}\ }\href
  {\doibase 10.1088/0022-3719/15/36/006} {\bibfield  {journal} {\bibinfo
  {journal} {J. Phys. C}\ }\textbf {\bibinfo {volume} {15}},\ \bibinfo {pages}
  {L1299} (\bibinfo {year} {1982})}\BibitemShut {NoStop}%
\bibitem [{\citenamefont {Widom}(1982)}]{Widom82}%
  \BibitemOpen
  \bibfield  {author} {\bibinfo {author} {\bibfnamefont {A.}~\bibnamefont
  {Widom}},\ }\bibfield  {title} {\enquote {\bibinfo {title} {{Thermodynamic
  derivation of the Hall effect current}},}\ }\href {\doibase
  10.1016/0375-9601(82)90401-7} {\bibfield  {journal} {\bibinfo  {journal}
  {Phys. Lett. A}\ }\textbf {\bibinfo {volume} {90}},\ \bibinfo {pages} {474}
  (\bibinfo {year} {1982})}\BibitemShut {NoStop}%
\bibitem [{\citenamefont {Thouless}\ \emph {et~al.}(1982)\citenamefont
  {Thouless}, \citenamefont {Kohmoto}, \citenamefont {Nightingale},\ and\
  \citenamefont {{den Nijs}}}]{Thouless82}%
  \BibitemOpen
  \bibfield  {author} {\bibinfo {author} {\bibfnamefont {D.~J.}\ \bibnamefont
  {Thouless}}, \bibinfo {author} {\bibfnamefont {M.}~\bibnamefont {Kohmoto}},
  \bibinfo {author} {\bibfnamefont {M.~P.}\ \bibnamefont {Nightingale}}, \ and\
  \bibinfo {author} {\bibfnamefont {M.}~\bibnamefont {{den Nijs}}},\ }\bibfield
   {title} {\enquote {\bibinfo {title} {{Quantized Hall Conductance in a
  Two-Dimensional Periodic Potential}},}\ }\href {\doibase
  10.1103/PhysRevLett.49.405} {\bibfield  {journal} {\bibinfo  {journal} {Phys.
  Rev. Lett.}\ }\textbf {\bibinfo {volume} {49}},\ \bibinfo {pages} {405}
  (\bibinfo {year} {1982})}\BibitemShut {NoStop}%
\bibitem [{\citenamefont {Niu}\ \emph {et~al.}(1985)\citenamefont {Niu},
  \citenamefont {Thouless},\ and\ \citenamefont {Wu}}]{Niu85}%
  \BibitemOpen
  \bibfield  {author} {\bibinfo {author} {\bibfnamefont {Q.}~\bibnamefont
  {Niu}}, \bibinfo {author} {\bibfnamefont {D.~J.}\ \bibnamefont {Thouless}}, \
  and\ \bibinfo {author} {\bibfnamefont {Y.~S.}\ \bibnamefont {Wu}},\
  }\bibfield  {title} {\enquote {\bibinfo {title} {{Quantized Hall conductance
  as a topological invariant}},}\ }\href {\doibase 10.1103/PhysRevB.31.3372}
  {\bibfield  {journal} {\bibinfo  {journal} {Phys. Rev. B}\ }\textbf {\bibinfo
  {volume} {31}},\ \bibinfo {pages} {3372} (\bibinfo {year}
  {1985})}\BibitemShut {NoStop}%
\bibitem [{Baa()}]{Baake_book00}%
  \BibitemOpen
  \href@noop {} {\enquote {\bibinfo {title} {{\it Directions in Mathematical
  Quasicrystals}},}\ }\bibinfo {note}
  {\href{http://bookstore.ams.org/crmm-13}{CRM Monograph Series Vol.13}, edited
  by M. Baake and R. V. Moody (AMS, Providence, 2000)}\BibitemShut {NoStop}%
\bibitem [{\citenamefont {Pannetier}\ \emph {et~al.}(1984)\citenamefont
  {Pannetier}, \citenamefont {Chaussy}, \citenamefont {Rammal},\ and\
  \citenamefont {Villegier}}]{Pannetier84}%
  \BibitemOpen
  \bibfield  {author} {\bibinfo {author} {\bibfnamefont {B.}~\bibnamefont
  {Pannetier}}, \bibinfo {author} {\bibfnamefont {J.}~\bibnamefont {Chaussy}},
  \bibinfo {author} {\bibfnamefont {R.}~\bibnamefont {Rammal}}, \ and\ \bibinfo
  {author} {\bibfnamefont {J.~C.}\ \bibnamefont {Villegier}},\ }\bibfield
  {title} {\enquote {\bibinfo {title} {{Experimental Fine Tuning of
  Frustration: Two-Dimensional Superconducting Network in a Magnetic Field}},}\
  }\href {\doibase 10.1103/PhysRevLett.53.1845} {\bibfield  {journal} {\bibinfo
   {journal} {Phys. Rev. Lett.}\ }\textbf {\bibinfo {volume} {53}},\ \bibinfo
  {pages} {1845} (\bibinfo {year} {1984})}\BibitemShut {NoStop}%
\bibitem [{\citenamefont {Goldman}\ \emph {et~al.}(2011)\citenamefont
  {Goldman}, \citenamefont {Urban},\ and\ \citenamefont
  {Bercioux}}]{Goldman11}%
  \BibitemOpen
  \bibfield  {author} {\bibinfo {author} {\bibfnamefont {N.}~\bibnamefont
  {Goldman}}, \bibinfo {author} {\bibfnamefont {D.~F.}\ \bibnamefont {Urban}},
  \ and\ \bibinfo {author} {\bibfnamefont {D.}~\bibnamefont {Bercioux}},\
  }\bibfield  {title} {\enquote {\bibinfo {title} {{Topological phases for
  fermionic cold atoms on the Lieb lattice}},}\ }\href {\doibase
  10.1103/PhysRevA.83.063601} {\bibfield  {journal} {\bibinfo  {journal} {Phys.
  Rev. A}\ }\textbf {\bibinfo {volume} {83}},\ \bibinfo {pages} {063601}
  (\bibinfo {year} {2011})}\BibitemShut {NoStop}%
\bibitem [{\citenamefont {Zak}(1964{\natexlab{b}})}]{Zak64}%
  \BibitemOpen
  \bibfield  {author} {\bibinfo {author} {\bibfnamefont {J.}~\bibnamefont
  {Zak}},\ }\bibfield  {title} {\enquote {\bibinfo {title} {{Group-Theoretical
  Consideration of Landau Level Broadening in Crystals}},}\ }\href {\doibase
  10.1103/PhysRev.136.A776} {\bibfield  {journal} {\bibinfo  {journal} {Phys.
  Rev.}\ }\textbf {\bibinfo {volume} {136}},\ \bibinfo {pages} {A776} (\bibinfo
  {year} {1964}{\natexlab{b}})}\BibitemShut {NoStop}%
\bibitem [{\citenamefont {Edwards}\ and\ \citenamefont
  {Thouless}(1972)}]{Edwards72}%
  \BibitemOpen
  \bibfield  {author} {\bibinfo {author} {\bibfnamefont {J.~T.}\ \bibnamefont
  {Edwards}}\ and\ \bibinfo {author} {\bibfnamefont {D.~J.}\ \bibnamefont
  {Thouless}},\ }\bibfield  {title} {\enquote {\bibinfo {title} {{Numerical
  studies of localization in disordered systems}},}\ }\href {\doibase
  10.1088/0022-3719/5/8/007} {\bibfield  {journal} {\bibinfo  {journal} {J.
  Phys. C}\ }\textbf {\bibinfo {volume} {5}},\ \bibinfo {pages} {807} (\bibinfo
  {year} {1972})}\BibitemShut {NoStop}%
\bibitem [{\citenamefont {Abrahams}\ \emph {et~al.}(1979)\citenamefont
  {Abrahams}, \citenamefont {Anderson}, \citenamefont {Licciardello},\ and\
  \citenamefont {Ramakrishnan}}]{Abrahams79}%
  \BibitemOpen
  \bibfield  {author} {\bibinfo {author} {\bibfnamefont {E.}~\bibnamefont
  {Abrahams}}, \bibinfo {author} {\bibfnamefont {P.~W.}\ \bibnamefont
  {Anderson}}, \bibinfo {author} {\bibfnamefont {D.~C.}\ \bibnamefont
  {Licciardello}}, \ and\ \bibinfo {author} {\bibfnamefont {T.~V.}\
  \bibnamefont {Ramakrishnan}},\ }\bibfield  {title} {\enquote {\bibinfo
  {title} {{Scaling Theory of Localization: Absence of Quantum Diffusion in Two
  Dimensions}},}\ }\href {\doibase 10.1103/PhysRevLett.42.673} {\bibfield
  {journal} {\bibinfo  {journal} {Phys. Rev. Lett.}\ }\textbf {\bibinfo
  {volume} {42}},\ \bibinfo {pages} {673} (\bibinfo {year} {1979})}\BibitemShut
  {NoStop}%
\bibitem [{\citenamefont {Werner}\ \emph {et~al.}(2015)\citenamefont {Werner},
  \citenamefont {Brataas}, \citenamefont {von Oppen},\ and\ \citenamefont
  {Zar\'and}}]{Werner15}%
  \BibitemOpen
  \bibfield  {author} {\bibinfo {author} {\bibfnamefont {M.~A.}\ \bibnamefont
  {Werner}}, \bibinfo {author} {\bibfnamefont {A.}~\bibnamefont {Brataas}},
  \bibinfo {author} {\bibfnamefont {F.}~\bibnamefont {von Oppen}}, \ and\
  \bibinfo {author} {\bibfnamefont {G.}~\bibnamefont {Zar\'and}},\ }\bibfield
  {title} {\enquote {\bibinfo {title} {{Anderson localization and quantum Hall
  effect: Numerical observation of two-parameter scaling}},}\ }\href {\doibase
  10.1103/PhysRevB.91.125418} {\bibfield  {journal} {\bibinfo  {journal} {Phys.
  Rev. B}\ }\textbf {\bibinfo {volume} {91}},\ \bibinfo {pages} {125418}
  (\bibinfo {year} {2015})}\BibitemShut {NoStop}%
\bibitem [{\citenamefont {Sutherland}(1986)}]{Sutherland86}%
  \BibitemOpen
  \bibfield  {author} {\bibinfo {author} {\bibfnamefont {B.}~\bibnamefont
  {Sutherland}},\ }\bibfield  {title} {\enquote {\bibinfo {title}
  {{Localization of electronic wave functions due to local topology}},}\ }\href
  {\doibase 10.1103/PhysRevB.34.5208} {\bibfield  {journal} {\bibinfo
  {journal} {Phys. Rev. B}\ }\textbf {\bibinfo {volume} {34}},\ \bibinfo
  {pages} {5208} (\bibinfo {year} {1986})}\BibitemShut {NoStop}%
\bibitem [{\citenamefont {Onsager}(1952)}]{Onsager52}%
  \BibitemOpen
  \bibfield  {author} {\bibinfo {author} {\bibfnamefont {L.}~\bibnamefont
  {Onsager}},\ }\bibfield  {title} {\enquote {\bibinfo {title} {{Interpretation
  of the de Haas-van Alphen effect}},}\ }\href {\doibase
  10.1080/14786440908521019} {\bibfield  {journal} {\bibinfo  {journal} {Phil.
  Mag.}\ }\textbf {\bibinfo {volume} {43}},\ \bibinfo {pages} {1006} (\bibinfo
  {year} {1952})}\BibitemShut {NoStop}%
\bibitem [{\citenamefont {Dirac}(1931)}]{Dirac31}%
  \BibitemOpen
  \bibfield  {author} {\bibinfo {author} {\bibfnamefont {P.~A.~M.}\
  \bibnamefont {Dirac}},\ }\bibfield  {title} {\enquote {\bibinfo {title}
  {{Quantised Singularities in the Electromagnetic Field}},}\ }\href {\doibase
  10.1098/rspa.1931.0130} {\bibfield  {journal} {\bibinfo  {journal} {Proc. R.
  Soc. London, Ser. A}\ }\textbf {\bibinfo {volume} {133}},\ \bibinfo {pages}
  {60} (\bibinfo {year} {1931})}\BibitemShut {NoStop}%
\bibitem [{\citenamefont {Dana}\ \emph {et~al.}(1985)\citenamefont {Dana},
  \citenamefont {Avron},\ and\ \citenamefont {Zak}}]{Dana85}%
  \BibitemOpen
  \bibfield  {author} {\bibinfo {author} {\bibfnamefont {I.}~\bibnamefont
  {Dana}}, \bibinfo {author} {\bibfnamefont {Y.}~\bibnamefont {Avron}}, \ and\
  \bibinfo {author} {\bibfnamefont {J.}~\bibnamefont {Zak}},\ }\bibfield
  {title} {\enquote {\bibinfo {title} {{Quantised Hall conductance in a perfect
  crystal}},}\ }\href {\doibase 10.1088/0022-3719/18/22/004} {\bibfield
  {journal} {\bibinfo  {journal} {J. Phys. C}\ }\textbf {\bibinfo {volume}
  {18}},\ \bibinfo {pages} {679} (\bibinfo {year} {1985})}\BibitemShut
  {NoStop}%
\end{thebibliography}

%

\end{document}